\documentclass[copyright]{eptcs}
\usepackage{amssymb}
\providecommand{\event}{DCFS 2010}

\def\newrmtheorem#1{\@ifnextchar[{\@ormthm{#1}}{\@nrmthm{#1}}}

\newtheorem{theorem}{\bf Theorem}[section]
\newtheorem{lemma}[theorem]{\bf Lemma}        
\newtheorem{proposition}[theorem]{\bf Proposition}  
\newtheorem{corollary}[theorem]{\bf Corollary}
\newtheorem{definition}[theorem]{\bf Definition}

\newtheorem{example}[theorem]{\bf Example}

\newcommand{\cA}{\mathcal{A}}
\newcommand{\cB}{\mathcal{B}}

\newcommand{\cD}{\mathcal{D}}
\newcommand{\zo}{\{0,1\}}
\newcommand{\zos}{\{0,1\}^*}
\newcommand{\lex}{<_{\mathrm{lex}}}
\newcommand{\mod}{\rm mod}
\newcommand{\ord}{{o}}
\newcommand{\pre}{\mathrm{pre}}

\def\Bbox{
{\unskip\nobreak\hfil\penalty50
\hskip1em\hbox{}\nobreak\hfil{\lower .5pt \hbox{$\Box$}}
\parfillskip=0pt \finalhyphendemerits=0 \par}
}

\def\eop{
\ifmmode {\hbox{\Bbox}} \else \Bbox \fi
}

\title{Representing Small Ordinals by Finite Automata}

\def\titlerunning{Small Ordinals and Finite Automata}
\def\authorrunning{Z. \'Esik}

\author{Z. \'Esik\thanks{Supported in part by grant no. K 75249 of the National Foundation of
Hungary for Scientific Research (OTKA) and the T\'AMOP 4.2.1/B program of the Hungarian National Development Agency.}
\institute{
Dept. of Computer Science\\
University of Szeged\\
Hungary}
\email{ze@inf.u-szeged.hu}
}

\begin{document}

\maketitle

\begin{abstract}
It is known that an ordinal is the order type of the lexicographic
ordering of a regular language if and only if it is less than
$\omega^\omega$. We design a polynomial time algorithm that 
constructs, for each  well-ordered regular language $L$
with respect to the lexicographic ordering, given by a deterministic 
finite automaton, the Cantor Normal Form of its order type. It follows that 
there is a polynomial time algorithm to decide whether two  
deterministic finite automata accepting well-ordered regular 
languages accept isomorphic languages.    
We also give estimates on the size of the smallest automaton
representing an ordinal less than $\omega^\omega$, together with
an algorithm that translates each such ordinal to an automaton.
\end{abstract}

\section{Introduction}

One of the basic decision problems in the theory of automata and languages
is the equivalence or equality problem that asks if two specifications
define equal languages. In this paper we study the related ``isomorphism problem''
of deciding whether the lexicographic orderings of the languages 
defined by two specifications are isomorphic, i.e., whether 
the two languages determine ``isomorphic dictionaries''.

The study of lexicographic orderings of regular languages, or equivalently, 
lexicographic orderings of the leaves of regular trees goes back
to \cite{Courcelle}. 
Thomas \cite{Thomas} has shown without giving any 
complexity bounds that it decidable whether the lexicographic orderings of two regular languages 
(given by finite automata or regular expressions) are isomorphic.
In contrast, the results in \cite{BEregwords} imply that there is an exponential 
algorithm to decide whether the lexicographic orderings 
of two regular languages, given by deterministic finite automata (DFA) 
are isomorphic. In contrast, no such algorithm exists
for context-free languages, cf. \cite{EsCF}.
In this paper, one of our aims is to show that there is a polynomial time 
algorithm to decide for DFA accepting lexicographically well-ordered 
languages, whether they accept isomorphic languages with respect
the lexicographic order.

The ordinals that arise as order types of lexicographic well-orderings 
of regular languages are exactly the ordinals less than $\omega^\omega$,
cf. \cite{BEord,Heilbrunner}. 
The Cantor Normal Form (CNF) \cite{Rosenstein} of any such nonzero ordinal 
takes the form
$\omega^{n_0}\times m_0 + \cdots + \omega^{n_k}\times m_k,$
where $k,n_i$ and $m_i$ are integers such that $k \geq 0$,
$m_i\geq 1$, $i = 0,\ldots,k$, and $n_0 >\cdots > n_k\geq 0$.
We provide an algorithm that, given an ``ordinal automaton'' 
representing a well-ordering, computes its CNF. 

We also give estimates on the size of the smallest ordinal automaton
representing an ordinal less than $\omega^\omega$, together with
an algorithm that translates such an ordinal to an automaton.

In the main part of the paper we will restrict ourselves to DFA over
the binary alphabet $\{0,1\}$ accepting a complete prefix language 
(complete prefix code). However, this restriction  
is only a technical convenience and is not essential for the results.

\section{Lexicographic orderings}

Suppose that $\Sigma$ is an alphabet linearly ordered by the relation $<$.
We define the lexicographic ordering $\lex$ of the set $\Sigma^*$ 
by $u \lex v$ iff $u$ is either a proper prefix of $v$ or $u$ and $v$ 
are of the form $u = xay$, $v = xbz$ with $a < b$ in $\Sigma$.
When $L\subseteq \Sigma^*$, we obtain a (strict) linear ordering 
$(L,\lex)$, called the \emph{lexicographic ordering of $L$}.
 It is known that if $\Sigma$ has two or more letters,
then every countable linear ordering is isomorphic to the
linear ordering $(L,\lex)$ of some language 
$L\subseteq \Sigma^*$, see e.g. \cite{BEord}. 
Moreover, we may restrict ourselves to prefix 
languages,
for if $L \subseteq \{a_1,\cdots,a_n\}$ where the 
alphabet is ordered as indicated, then  $(L,\lex)$ is isomorphic to $(La_0,\lex)$,
where $a_0$ is a new letter which is  lexicographically less than any other letter.
Further, we may restrict ourselves to the binary alphabet, 
since each ordered alphabet of $n$ letters can be encoded by words 
over $\zo$ of  length $\lceil \log n \rceil$ in an order preserving 
manner. 
Actually, it suffices to consider 
\emph{complete} prefix languages $L \subseteq \zos$ having the 
property that for any $u \in \zos$, $u0$ is in the set $\pre(L)$ 
of all prefixes of words in $L$ iff $u1 \in \pre(L)$. 

Suppose that $L\subseteq \zos$ is a complete prefix language.
We define the complete binary tree $T_L$ to be the tree whose 
vertices are the words in $\pre(L)$, 
such that each vertex $u\in \pre(L)$ is either a leaf or has 
two successors, the words $u0$ and $u1$. When $L$ is the 
empty language, $T_L$ is the empty tree. Note that $T_L$ 
is an ordered tree, since the successors $u0,u1$ 
of a non-leaf vertex $u$ are ordered by $u0 \lex u1$. 
The linear ordering $(L,\lex)$ is just the ordering 
of the leaves of $T_L$. Note that each infinite branch
of $T_L$ determines an $\omega$-word over $\{0,1\}$. 
Below we will make use of the following 
simple fact, see also \cite{Carayoletal}. 

\begin{lemma}
\label{lem-intro}
Suppose that $L \subseteq \zos$ and consider the tree $T_L$.
Then $(L,\lex)$ is a well-ordering iff the $\omega$-word 
determined by each infinite branch of $T_L$ contains a 
finite number of occurrences of $0$.
\end{lemma}

Call a linear ordering regular if it is isomorphic to the 
lexicographic ordering of a regular (complete prefix) language over some 
ordered alphabet, or equivalently, over the alphabet $\zo$.
A \emph{regular well-ordering} is a regular linear ordering that is 
a well-ordering. 

Regarding linear orderings and ordinals, we will use standard 
terminology. Below we review some simple facts for 
linear orderings and ordinal arithmetic
(restricted to ordinals less than $\omega^\omega$). For all 
unexplained notions we refer to \cite{Rosenstein}.

Suppose that $P = (P,<_P)$ and $Q= (Q,<_Q)$ are disjoint 
(strict) linear orderings.  
Then the \emph{ordered sum} $P + Q$ is the linear ordering 
$(P \cup Q, <)$, where the restriction of $<$ to $P$ is the relation $<_P$
and similarly for $Q$, and where $x < y$ holds for all $x \in P$ and $y \in Q$.
It is known that if $P$ and $Q$ are well-ordered of order type
$\alpha$ and $\beta$, respectively, where $\alpha$ and $\beta$ are ordinals, 
then $P+Q$ is well-ordered of order type $\alpha + \beta$. 
In addition to sum, we will make use of the product operation. 
Given $P$ and $Q$ as above, let us define the following 
linear order $<$ of the set $P \times Q$: For all $(x,y),(x',y') \in P \times Q$, 
$(x,y) < (x',y')$ iff $y <_Q y'$, or $y = y'$ and $x<_P x'$. 
When $P,Q$ are well-ordered of order type $\alpha,\beta$, respectively, then 
$P \times Q$ is also well-ordered of order type $\alpha \times \beta$. 

As mentioned in the Introduction, it is known that a well-ordering 
is regular iff its order type is less than the ordinal $\omega^\omega$.
The Cantor Normal Form (CNF) \cite{Rosenstein} 
of each nonzero ordinal less than this bound is of the form
$
\omega^{n_0}\times m_0 + \cdots + \omega^{n_k}\times m_k,
$
where $k \geq 0$ and $n_i$ and $m_i$ are integers with  
$n_0 > \cdots > n_k \geq 0$, $m_i \geq 1$ for all $i = 0,\cdots,k$. The 
exponent $n_0$ is called the \emph{degree}.  

In order to compute the CNF of the 
sum of two nonzero ordinals less than $\omega^\omega$, it is helpful to know 
that $\omega^m +\omega^n = \omega^n$ whenever $m < n$. 
Thus, when 
\begin{eqnarray*}
\alpha = \omega^{n_0}\times m_0 + \cdots + \omega^{n_k}\times m_k
\quad {\rm and}\quad 
\beta =  \omega^{n'_0}\times m'_0 + \cdots + \omega^{n'_\ell}\times m'_\ell,
\end{eqnarray*}
then the CNF of $\alpha + \beta$ can be computed as follows. 
First, suppose that $n_0,\cdots,n_{i-1}$ are all greater than $n'_0$
and $n_{i} \leq n'_0$. If $n_i = n'_0$, then $\alpha + \beta$
is
$$\omega^{n_0}\times m_0 + \cdots + \omega^{n_{i-1}}\times m_{i-1} 
+ \omega^{n_i}\times (m_i + m'_0) +
 \omega^{n'_1}\times m'_1 + \cdots + \omega^{n'_\ell}\times m'_\ell.$$
If $n_i < n'_0$, then $\alpha + \beta$ is
$$
\omega^{n_0}\times m_0 + \cdots + \omega^{n_{i-1}}\times m_{i-1}
+  \omega^{n'_0}\times m'_0 + \cdots + \omega^{n'_\ell}\times m'_\ell.
$$
Finally, suppose that $n_k > n'_0$. In that case $\alpha + \beta$ is 
$$
\omega^{n_0}\times m_0 + \cdots + \omega^{n_k}\times m_k + 
 \omega^{n'_0}\times m'_0 + \cdots + \omega^{n'_\ell}\times m'_\ell.
$$
In order to compute the product $\alpha \times \beta$, it suffices to know 
that product distributes over sum on the left, and if $\alpha$ is 
the ordinal given above, then $\alpha \times \omega = \omega^{n_0 +1}$.



\section{Ordinal automata}

We will be considering DFA $\cA = (Q,\zo,\delta,q_0,F)$,
where $Q$ is the finite set of states, $\zo$ is the
input alphabet, $\delta$ is a partial function
$Q \times \zo \to Q$, the transition function,
$q_0\in Q$ is the initial state, and $F \subseteq Q$ 
is the set of final states. As usual, we extend $\delta$ 
to a partial function $Q \times \zos \to Q$ 
and write $qu$ for $\delta(q,u)$, for $q \in Q$ and
$u \in \zos$.

The \emph{language $L(\cA)$ accepted by the DFA} 
$\cA= (Q,\zo,\delta,q_0,F)$ is the set $\{ u \in \zos: 
q_0u \in F\}$. 
As usual, we call an automaton $\cA = (Q,\zo,\delta,q_0,F)$ \emph{trim}
if each state $q\in Q$ is both accessible and
co-accessible, i.e., when there exist words $u,v \in
\zos$ with $q_0u = q$ and $qv \in F$. It is well-known that 
if $L(\cA)$ is nonempty, then $\cA$ is equivalent to a 
trim automaton that can be easily constructed from $\cA$ 
by removing all states that are not accessible or co-accessible.
\emph{To avoid trivial situations, we will only consider 
automata that accept a nonempty language, so that we 
may restrict ourselves to trim automata}.

A trim automaton $\cA = (Q,\zo,\delta,q_0,F)$
accepts a prefix language iff neither $q0$ nor $q1$ is defined 
when $q \in F$.  Moreover, assuming that 
this holds, $\cA$ accepts a complete prefix language iff
for every $q\in Q\setminus F$, both $q0$ 
and $q1$ are defined. We will call such trim 
automata \emph{complete prefix automata} (CPA). It is clear that 
for each trim automaton $\cA$ accepting a prefix language 
one can construct a CPA $\cA' = (Q',\zo,\delta',q_0,F')$ 
with $Q' \subseteq Q$  such that $(L(\cA),\lex)$ is isomorphic to 
$(L(\cA'),\lex)$. To this end, for each state $q\in Q$ we 
form the unique sequence of states $q = q_1,q_2,\ldots,q_k$ 
such that for each $1 \leq i < k$, $q_{i+1} = q_i0$ or $q_{i+1} = q_i1$, 
moreover, exactly one of $q_i0$ and $q_i1$ is defined, and finally
either $q_k \in F$ (in which case neither $q_k0$ nor $q_k1$ is defined), 
or both $q_k0$ and $q_k1$ are defined. If $q_k \in F$, then we remove 
the transitions used to form this sequence and declare $q$ to be a final state. 
If $q_k \not\in F$, then we replace the transition originating in $q$
by the two transitions $\delta'(q,i) = \delta(q_k,i)$, $ i = 0,1$.
Finally, we remove states that are not accessible or co-accessible.

Suppose that  $\cA = (Q,\zo,\delta,q_0,F)$ is a DFA. By the \emph{size} of $\cA$ 
we will mean the number of states in $Q$. 
The strongly connected components of $\cA$ 
are defined as usual. We say that a strongly connected component $C$ is \emph{trivial} 
if $C$ consists of a single state $q$ and $q \not\in \{q0,q1\}$.
Otherwise $C$ is called \emph{nontrivial}. 
We impose  the usual  
partial order on strongly connected components by defining $C \preceq C'$ iff 
there exist some $q \in C$ and $u \in \zos$ with $qu \in C'$. The \emph{height}
of a nontrivial strongly connected component $C$ is the length $k$ of the 
longest sequence $C_1,\ldots,C_k$ of nontrivial strongly connected 
components such that $C_1\prec \cdots \prec C_k$ and $C_k = C$. 
From Lemma~\ref{lem-intro} we immediately have:

\begin{proposition}
A CPA $\cA = (Q,\zo,\delta,q_0,F)$ accepts a well-ordered 
language iff for each nontrivial strongly connected component
$C$ and $q \in C$ it holds that $q0 \not\in  C$ (and
of course $q1 \in C$).
\end{proposition}

We conclude that there is a simple algorithm to decide 
whether a CPA  accepts a well-ordered language which 
runs in polynomial time in the size of the automaton,
see also \cite{BEscattered,BloomZhang}.   
It is trivial to extend this result to automata over 
larger alphabets.

\begin{definition}
\label{def-ordinal aut}
An \emph{ordinal automaton} (OA) is a CPA 
$\cA = (Q,\zo,\delta,q_0,F)$ 
such that whenever $q$ belongs to a nontrivial
strongly connected component $C$, $q0$ does not
belong to $C$.
\end{definition}

By the previous proposition, a CPA $\cA$ is an OA
iff it accepts a well-ordered (complete prefix) 
language. For an OA $\cA$, we call the order type of $(L(\cA),\lex)$ 
the \emph{ordinal represented by $\cA$}, denoted $\ord(\cA)$. 

\begin{lemma}
\label{lem-omega n}
For each $n\geq 0$, there is an OA $\cA_n$ of size $n+1$ representing $\omega^n$.
\end{lemma} 

{\sl Proof.}\  Let $\cA_n$ have states $s_0,\cdots,s_n$ with transitions 
$\delta(s_i,1) = s_i$ and $\delta(s_i,0) = s_{i-1}$ for all $1 \leq i \leq n$.
The initial state is $s_n$ and the only final state is $s_0$. \eop 

\begin{example}
Consider the ordinal $\alpha = \omega^3 \times 2 + \omega$. 
An ordinal automaton representing $\alpha$ has $6$ states, $q_0,q_1,s_0,s_1,s_2,s_3$,
where $q_0$ is the initial state and $s_0$ is the only final state. 
The transitions are defined by $q_00 = q_1$, $q_0 1 = s_1$, 
$q_10 = q_11 = s_3$, and $s_i 1 = s_i$, $s_i0 = s_{i-1}$ for 
$1 = 1,2,3$. 
\end{example} 

We end this section with a construction converting a nonzero ordinal $\alpha < \omega^\omega$
to an OA. First, for each $n\geq 1$, we construct a CPA $\cD_n$ 
having a single final state which accepts a language of $n$ words.
 The CPA $\cD_1$ has a single state which is both 
initial and final, and no  transitions.  
If $n$ is even, say $n = 2k$, consider  $\cD_k$ and add a new initial state $s_0$ 
together with  transitions $s_00 = s_01 = s_0'$ to the old initial state $s_0'$. 
The only final state is the final state of $\cD_k$. 
If $n = 2k + 1$ for some $k\geq 1$, then consider $\cD_k$ with initial state $s_0'$ 
and final state $s_f$. We add two new states $s_0$ and $s_1$ and new transitions 
$s_00 = s_1$, $s_01 = s_f$, $s_10 = s_11=  s_0'$. 

Now let the CNF of $\alpha$ be $\omega^{n_0}\times m_0 + \cdots + 
\omega^{n_k}\times m_k$. When $k = 0$ and $m_0 =1$, then 
we may take the OA $\cA_{n_0}$ of Lemma~\ref{lem-omega n},
we have that $\ord(\cA_{n_0}) = \alpha$. 
So suppose that $k > 0$ or $m_0 > 1$. For each $0 \leq i \leq k$, consider the 
automaton $\cD_{m_i}$ constructed above
with initial state $q_i$ and final state $c_i$, say. We may assume 
that the state sets of the $\cD_{m_i}$ are pairwise disjoint. 
Then we form the ``ordered sum'' of the $\cD_{m_i}$, $i = 0,\cdots,k$ by adding 
$k$ new states $s_0,\cdots,s_{k-1}$, transitions $s_01 = s_1, \cdots, s_{k-2}1 = s_{k-1}$,
$s_0 0 = q_0, \cdots,s_{k-2}0 = q_{k-2}$, $s_{k-1}0 = q_{k-1}$ and $s_{k-1}1 = q_k$.  
Finally, take the automaton $\cA_{n_0}$ 
of Lemma~\ref{lem-omega n}, and identify its state $s_{n_i}$ with $c_i$ for all 
$i = 0,\cdots,k$. 
The resulting OA has $n_0 + g(m_0) + \cdots + g(m_k)$ states and represents $\alpha$,
where 
$g(1) = 1$ and  
$g(2m) = 1 + g(m)$, 
$g(2m + 1) = 2 + g(m)$
for all $m \geq 1$.

\section{From ordinal automata to CNF}

For this section, fix an OA 
$\cA = (Q,\zo,\delta,q_0,F)$. 
For each $q \in Q$, let us denote by $\cA_q$ the automaton 
$(Q_q,\zo,\delta_q,q,F_q)$, where $Q_q = \{qu : u\in\zos\}$,
$\delta_q$ is the restriction of $\delta$ to $Q_q \times \zo$,
and $F_q = Q_q \cap F$.

The following lemma is clear.

\begin{lemma}
For each state $q$, $\cA_q = (Q_q,\zo,\delta_q,q,F_q)$ is also 
an ordinal automaton.
\end{lemma}

For each $q \in Q$, we let $\ord(q)$ denote the order type of  
$(L_q,\lex) = (L(\cA_q),\lex)$. By the above lemma, $\ord(q)$ is a (nonzero) 
ordinal for each $q \in Q$.

\begin{lemma}
\label{lem-mon}
For all $q\in Q$ and $u\in \zos$, 
$\ord(qu) \leq \ord(q)$ Thus, if $q$
and $q'$ belong to the same strongly 
connected component, then $\ord(q) = \ord(q')$. 
\end{lemma} 

{\sl Proof.}\  The function $v \mapsto uv$, $v \in \zos$ defines an order embedding 
of the linear ordering $(L_{qu},\lex)$ into $(L_q,\lex)$. \eop

\begin{proposition}
\label{prop-strongly connected}
If $C$ is a \emph{nontrivial} strongly connected component,
then there is an integer $n \geq 1$ such that 
for all $q \in C$ it holds that $\ord(q) = \omega^n$.
Moreover, for each $q \in C$ the degree of $\ord(q0)$ is at most $n-1$, 
and there is some state $q'\in C$ such that the degree of 
$\ord(q'0)$ is $n -1$. 
\end{proposition}

{\sl Proof.}\  
By Definition~\ref{def-ordinal aut}, we can arrange the states in $C$ 
in a sequence $s_0,\cdots,s_{k-1}$ such that $s_i1 = s_{i+1\ \mod\ k}$ 
for all $i$. We also know that $s_i0 \not \in C$ for all $i$. Thus,
\begin{eqnarray}
\label{eq-sc}
\ord(s_0) = (\ord(s_00)+\cdots + \ord(s_{k-1}0))\times \omega = \alpha \times \omega.
\end{eqnarray} 
Since $0 < \alpha < \omega^\omega$, this is possible only if $\ord(s_0) = \omega^n$ 
for some $n\geq 1$. It follows now by Lemma~\ref{lem-mon} that 
$\ord(s_i) = \omega^n$ for all $i$. Using the formula (\ref{eq-sc}), 
it follows that the degree of each $\ord(s_i0)$ is at most $n-1$. 
Moreover, there is at least one $i_0$ such that the degree of 
$\ord(s_{i_0}0)$ is exactly $n-1$, since otherwise the degree
of $\ord(s_0)$ would be less than $n$. \eop 

When $C$ is a strongly connected component, trivial or not, 
we let $\ord(C)$ denote the ordinal $\ord(q)$ for $q \in C$. 

Suppose that the strongly connected component containing $q$ is trivial. 
\emph{Below we will say that a word $u$ {\bf leads} from $q$ to a 
strongly connected component $C$} if $qu \in C$ 
but $qv$ does not belong to any nontrivial strongly connected 
component whenever $v$ is a proper prefix of $u$.
When $C$ is a trivial strongly connected component consisting
of a single final state $q'$, then we also say that $u$ leads from
$q$ to the final state $q'$. The following fact is clear.

\begin{proposition}
\label{prop-sum}
Suppose that the strongly connected component of the state $q$ is trivial. 
Then let $u_1,\ldots,u_k$ denote in lexicographic order 
all the words leading from $q$ to a nontrivial strongly 
connected component, or to a final state.\footnote{The number of such 
words is clearly finite.} Then 
$\ord(q) = \ord(qu_1) + \cdots + \ord(qu_k).$
Thus, the degree of $\ord(q)$ is the maximum degree of the ordinals 
$\ord(qu_i)$, $i = 1,\cdots,k$.
\end{proposition}

We now prove a stronger version of Proposition~\ref{prop-strongly connected}. 

\begin{proposition}
\label{prop-strongly connected 2}
If $C$ is a nontrivial strongly connected component of  height $n$, 
then $\ord(C) = \omega^n$.
\end{proposition}

{\sl Proof.}\  
Suppose that $C$ is a nontrivial strongly connected 
component of height $n$. Clearly, $n \geq 1$. We argue by induction on $n$ 
to prove that $\ord(C) \geq \omega^n$. This is clear when 
$n = 1$, since by Proposition~\ref{prop-strongly connected}, 
$\ord(C) = \omega^m$ for some $m > 0$. Suppose now
that $n > 1$. Then let $C'$ be a nontrivial strongly 
connected component of height $n - 1$ accessible from a state 
of $C$ by some word. Then there exists a state $q \in C$ 
with $q0 \not\in C$ such that $C'$ is accessible from $q0$
by some word. Since $\ord(q0) \geq \ord(C') \geq \omega^{n-1}$, the degree of 
$\ord(q0)$ is at least $n - 1$. By (\ref{eq-sc}),  
$\ord(C) \geq \omega^n$.


Next we show that for any nontrivial strongly connected component $C$ of height $n$,
$\ord(C) \leq \omega^n$. This is clear when $n = 0$. Supposing $n > 0$, 
by Propositions~\ref{prop-sum} and the 
induction hypothesis we know that the degree
of $\ord(s0)$ is at most $n -1$ for each $s \in C$. Thus, by 
Proposition~\ref{prop-strongly connected}, $\ord(C) \leq \omega^n$. 
\eop

\begin{corollary}
\label{cor-omega n}
If the degree of $\ord(\cA)$ is $n$, then $\cA$ has at least $n+1$ states. 
\end{corollary}

{\sl Proof.}\  
This is clear when $n = 0$. Suppose now that $n > 0$. 
Since the degree of $\ord(\cA)$ is $n$, $\cA$ has at least 
one nontrivial strongly connected component of height $n$,
and thus at least one nontrivial strongly connected component of height $i$ for 
every $1 \leq i \leq n$.  Together with a final state, this gives at least $n + 1$ states. \eop


As a corollary of the above facts, there is an algorithm 
that computes the CNF of the ordinal $\ord(\cA)$ represented
by the ordinal automaton $\cA$. First, using some  standard polynomial time
algorithm,  we determine the set $K$  
of all nontrivial strongly connected components together with all 
trivial strongly connected components consisting of 
a single final state. We also determine $\ord(C) = \omega^n$ for each nontrivial 
strongly  connected component $C \in K$
by computing the height $n$ of $C$. We set $\ord(C) = 1$
for all strongly connected components $C \in K$ consisting
of a single final state.  If the initial state belongs to
some $C\in K$, then $\ord(\cA) = \ord(C)$. 
Otherwise let $n$ denote the maximum of the heights of the 
nontrivial strongly connected components, and let $n = 0$ if there 
is no nontrivial strongly connected component. Let $K_n$ denote the set 
of all nontrivial strongly connected components in $K$
of maximum height $n$. 
Using the algorithms specified in the Appendix as subroutines
with suitable parameters,
we determine for each $C\in K_n$ the number $m_C$ of all words $u$ 
leading from the initial state $q_0$ to $C$, together with the 
lexicographically greatest such word $u_C$. Then we define $x_n$ as the lexicographically 
greatest word among the $u_C$ and $m_n = \sum_{C \in K_n} m_C$.
By Proposition~\ref{prop-strongly connected 2}
and Proposition~\ref{prop-sum}, $\ord(\cA) = \omega^n \times m_n + \alpha_{n-1}$ for some 
unknown ordinal $\alpha_{n-1}$ of degree $n-1$. 

In the next step, we consider the set $K_{n-1}$ of all 
strongly connected components $C$ in $K$ of height $n-1$, 
and for each $C\in K_{n-1}$, we compute the number $m_C$ of all those words 
leading from $q_0$ to $C$ that are \emph{lexicographically greater
than} $x_n$, together with the lexicographically greatest such word
$u_C$, if any. Then $\alpha_{n-1} = \omega^{n-1} \times m_{n-1} + \alpha_{n-2}$, 
where $m_{n-1}$ is the sum of the integers $m_C$, $C \in K_{n-1}$,
 and $\alpha_{n-2}$ is some unknown ordinal of degree $n-2$. 
We also determine the lexicographically greatest 
word in the set consisting of $x_n$ and all words $u_C$, $C \in K_{n-1}$ 
such that $m_C > 0$, and we denote this word by $x_{n-1}$. 

Repeating the procedure, before the last step we know that 
$\ord(\cA) = \omega^n\times m_n  + \cdots + \omega \times m_1  +  \alpha_0$
where $\alpha_0 = m_0$ is an unknown finite ordinal. Moreover, we have computed a word 
$x_1$. In the last step, we consider the set $K_0$ of those connected
components in $K$ that consist of a single final state. We determine
for each $C \in K_0$ the number of all words leading from $q_0$ 
to $C$ that are lexicographically greater than $x_1$. Then 
$m_0 = \sum_{C \in K_0}m_C$.

We conclude that $\ord(\cA) = \omega^n \times m_n + \cdots + \omega \times m_1 + m_0$.
To get the CNF of $\alpha$, we remove all summands $\omega^i \times m_i$ with $m_i = 0$. 

The length of each word $u_C$ determined in the above algorithm is 
bounded by the size of $\cA$ and can be determined in 
polynomial time. Similarly, the length
of the binary representation of each $m_C$ is at most the size of $\cA$,
and each $m_C$ can be computed in polynomial time in the size of $\cA$.
Thus, the overall algorithm runs in polynomial time.  
We have proved:

\begin{theorem}
There is a polynomial algorithm that, given an ordinal automaton
$\cA$, computes the CNF of the ordinal $\ord(\cA)$ represented by $\cA$.
\end{theorem}

\begin{corollary}
There is a polynomial time algorithm to decide for 
ordinal automata $\cA$ and $\cB$ whether 
$\ord(\cA) = \ord(\cB)$, i.e., whether $(L(\cA),\lex)$
and $(L(\cB),\lex)$ are isomorphic.
\end{corollary}

{\sl Proof.}\  We compute in polynomial time the CNFs of $\ord(\cA)$ and $\ord(\cB)$
and check whether they are identical. \eop

\section{Minimal ordinal automata}

For a nonzero ordinal $\alpha < \omega^\omega$, let $\#(\alpha)$ 
denote the minimum number $m$ such that $\alpha = \ord(\cA)$ for 
some $m$-state OA $\cA$. In this section we reduce the determination
of the function $\#(\alpha)$ to another problem on automata
and give some estimations on $\#(\alpha)$ in terms of the 
CNF of $\alpha$. 

\begin{definition}
\label{def-f}
Let $m_0,\cdots,m_k$ be positive integers. 
Then we let 
$f(m_0,\cdots,m_k)$ 
denote the minimal number of states of a CPA
$\cA = (Q,\zo,\delta,q_0,F)$ having no  nontrivial strongly connected component 
with the following properties:
\begin{enumerate}
\item $F = \{c_0 ,\cdots, c_k\}$, where the states $c_i$ are pairwise different.
\item For each $i$ with $0 \leq i \leq k$, the language $L_i$ has exactly $m_i$ words: 
$$L_i = \{u \in \zos: q_0u = c_i\ \wedge \ \forall v,j\ ((u \lex v\ \wedge\ q_0v = c_j)\ 
            \Rightarrow j \geq i)\}$$
\end{enumerate}
\end{definition} 
Note that $f(m_0,\cdots,m_k) \geq k + 1$. 
Also, when $m \geq 1$, $f(m)$ is the minimum number of states of a CPA 
accepting a language of $m$ words. In particular, $f(1) = 1$.

\begin{lemma}
\label{lem-minoa}
Suppose that $\alpha$ is a nonzero ordinal 
with CNF  $\omega^{n_0}\times m_0 + \cdots + \omega^{n_k} \times m_k$. 
Then there is a OA of size $n_0 - k + f(m_0,\cdots,m_k)$ 
representing $\alpha$.   
\end{lemma}

{\sl Proof.}\  We take a CPA $\cA = (Q,\zo,\delta,q_0,F)$ having no nontrivial strongly connected 
component as in Definition~\ref{def-f}, having $f(m_0,\cdots,m_k)$ states, 
 and the automaton $\cA_{n_0}$ constructed 
in Lemma~\ref{lem-omega n}. Then we identify $c_i$ with $s_{n_i}$ for all 
$i = 0,\cdots,k$.
\eop


\begin{theorem}
Suppose that the Cantor normal form of a nonzero ordinal $\alpha< \omega^\omega$ is 
$\omega^{n_0}\times m_0 + \cdots + \omega ^{n_k}\times m_k$. 
Then $\#(\alpha) = n_0 - k + m$ where $m = f(m_0,\cdots,m_k)$.
\end{theorem}

{\sl Proof.}\  
We have already shown that $\#(\alpha) \leq  n_0 - k + m$. Thus, it 
remains to prove that $\#(\alpha) \geq  n_0 - k + m$.

Suppose that $\cA = (Q,\zo,\delta,q_0,F)$ is an OA with $\ord(\cA) = \alpha$ having a 
least number of states among all such automata.
By Corollary~\ref{cor-omega n}, $\cA$ must have at least $n_0+1$ states. 
Thus, when $k = 0$ and $m_0 =1$, $\#(\alpha) \geq  n_0 + 1 = n_0 -k + m$, 
since $f(1) = 1$. So from now on we assume that $k > 0$ or $m_0 > 1$.

By the proof of Corollary~\ref{cor-omega n}, $\cA$ has at least one nontrivial 
strongly connected component  of height $i$ for each $i$ with $1 \leq i \leq n_0$, 
and of course  at least one final state. 
It is not possible that a nontrivial strongly connected 
component $C$ of height $n$, say,  contains two or more states, since otherwise 
we could select a state $q$ of $C$ such that at least 
one strongly connected component $C'$ of height $n-1$ is accessible 
from $q0$ by some word $u$ (i.e., $q0u \in C'$), and redirect any transition going to $C$ to 
the selected  state $q$. After that, 
we could remove all states in $C \setminus \{q\}$, the resulting ordinal 
automaton would still represent the same ordinal, by 
Proposition~\ref{prop-sum} and Proposition~\ref{prop-strongly connected 2}. 
Similarly, for each $1 \leq i \leq n_0$, there must be a single nontrivial
strongly connected component of height $i$. Indeed, if $C$ and $C'$ 
were different nontrivial strongly connected components of the 
same height $i$, then we could remove $C'$ and redirect every transition 
originally going to some state in $C'$ to a state in $C$; 
the resulting smaller OA would represent the 
same ordinal. Clearly, $\cA$  has a single final state. 
Also, if a state $q$ forms a nontrivial strongly connected component 
of height $i$, and $q'$ is either the state that forms the single 
nontrivial strongly connected component of height $i-1$ if $i>1$ 
or $q'$ is the single final state if $i =0$, and if $q0$ is not $q'$,
then we can redirect this transition from $q$ under $0$ to $q'$. The OA obtained after removing 
those states that possibly become inaccessible from the initial state 
 still represents $\alpha$. 

In conclusion, we have that $\cA$ contains a subautomaton consisting 
of states $s_{n_0},\cdots,s_0$ such that $s_{0}$ is the final state
and for each $i \geq 1$, $s_i$ forms a nontrivial strongly connected component of height $i$.
Moreover, $s_i 1 = s_i$ and $s_i 0 = s_{i-1}$ for all $i\geq  1$. 
Let $S = \{s_0,\cdots,s_{n_0}\}$. 
None of the states in $Q\setminus S$ is contained in any nontrivial strongly
connected component, and each state is accessible from $q_0$
by some word. Moreover, from each state $q \in Q\setminus S$ there is at least 
one word leading to some connected component $\{s_i\}$, 
trivial or not. We claim that if $q0 = s_i$ or $q1 = s_i$ for some $q \in Q\setminus S$ 
and $0\leq i \leq n_0$, then there exists some $0 \leq j \leq k$ 
with $i = n_j$, i.e., $\omega^i$ appears in the CNF of $\alpha$.
Indeed, if $q0 = s_i$, say,  but $i$ is not in the set $\{n_0,\cdots,n_k\}$,
then we can remove state $q$ and redirect all transitions going to 
$q$ to $q1$, the resulting smaller OA still represents $\alpha$,
a contradiction.

Since $k > 0$ or $m_0 > 1$,  the initial state $q_0$ is not in $S$
(since otherwise $\ord(q_0) = \ord(\cA)$ would be a power of $\omega$). 
Let us order the set $U$ of all words leading from $q_0$ to 
a strongly connected component $\{s_i\}$, $i = 0,\cdots,n_0$ 
lexicographically. We know that 
for each $u\in U$, $q_0u \in S' = \{s_{n_j} : 0 \leq j \leq k\}$. 
Then, by Proposition~\ref{prop-sum}, in order to have $\ord(q_0) = \alpha$, 
for each $j$ with $0 \leq j \leq k$ there must be exactly $m_j$ 
words $u \in U$ with $q_0u = s_{n_j}$ and such that 
there is no lexicographically greater word $v \in U$ with  
$q_0v \in \{s_{n_0},\cdots,s_{n_{j-1}}\}$. This means that 
by removing all states in $S \setminus S'$ and all transitions 
originating in the states belonging $S'$, the resulting 
automaton has at least $f(m_0,\cdots,m_k)$ states, and thus 
$\cA$ has at least $n_0 - k + m$ states. \eop

\begin{corollary}
For each $n\geq 0$, there is up to isomorphism a unique OA with $n+1$ states representing $\omega^n$, 
the automaton $\cA_n$ constructed in Lemma~\ref{lem-omega n}. 
\end{corollary}

\subsection{The function $f$}
\label{sec-comb}

In this section, we give some estimations on the function $f$ introduced 
above. 

\begin{proposition}
For all positive integers $m_0,\cdots,m_k$, 
$$f(m_0 + \cdots + m_k) \leq f(m_0,\cdots,m_k) \leq f(m_0) + \cdots + f(m_k) +k $$
\end{proposition}

{\sl Proof.}\  This is clear when $k = 0$, so assume that $k > 0$. 
To prove the upper bound, for each $m_i$ consider a CPA $\cB_i$ 
of size $f(m_i)$ without nontrivial strongly connected components 
and  having a single final state $c_i$
 which accepts a language 
of $m_i$ words. Without loss of generality, we may assume that the 
sets $Q_i$ are pairwise disjoint. 
Let $\cB$ be the ordered sum of the $\cB_i$ constructed as above. 
Then for each $i$, there are exactly $m_i$ words taking the initial state $s_0$ 
to $c_i$, and whenever $s_0u = c_i$ and $s_0v = c_j$ with $u \lex v$, 
it holds that $i \leq j$. Since $\cB$ has $f(m_0) + \cdots + f(m_k) +k$
states, we conclude that $f(m_0,\cdots,m_k) \leq f(m_0) + \cdots + f(m_k) +k$. 

To prove the lower bound, consider the automaton $\cB'$ obtained from $\cB$
by collapsing the final states $c_0,\cdots,c_k$ into a single final state. 
Then $\cB'$ accepts a language of $m_0 + \cdots + m_k$ words
and has $f(m_0,\cdots, m_k)$ states. Thus, 
$f(m_0 + \cdots + m_k) \leq f(m_0,\cdots,m_k)$. \eop

In the rest of this section, we consider the case when $k = 0$.
In this case, $f$ is a function on the positive integers. It is
not difficult to see that for each $n > 0$, 
$f(n)$ is the length of the shortest \emph{addition chain} \cite{Guy} 
representing $n$, 
i.e., $f(n)$ is the least integer $k$ for which there there exist 
different integers $1 = a_1 < \cdots < a_k = n$ such that for each 
$i > 1$ there exist $j_1,j_2$ with $a_i = a_{j_1} + a_{j_k}$.
Addition chains have a vast literature
\cite{Sloane} . It is not difficult to show that
$f(n)$ is at most the sum of $\log n$ and the number $m$ of occurrences 
of the digit $1$ in the binary representation of $n$. If $n$ is a power 
of $2$, then $f(n) = \log n$. 
In the first paper \cite{Schonhage} published in the journal TCS, it was shown that $f(n)$ 
is at least $\log n + \log m - 2.13$, where $m$ is defined as above.  
By \cite{Downeyetal}, it is an NP-complete problem to decide for 
integers $n,k \geq 1$ whether $f(n) \leq k$ holds.

\section{Conclusion and open problems}

We have shown that there is a polynomial time algorithm
to decide if two ordinal automata represent the same ordinal.
Since it is decidable in polynomial time whether the 
lexicographic ordering of the language accepted by a 
DFA is well-ordered, and since every DFA accepting a 
well-ordered regular language can be transformed in
polynomial time to an ordinal automaton, the restriction 
to ordinal automata was inessential.

A linear ordering is called \emph{scattered} if it does not have
a subordering isomorphic to the dense ordering of the rationals.
By Hausdorff's theorem \cite{Rosenstein}, every linear ordering is a dense 
sum of scattered linear orderings. Call a language 
scattered if its lexicographic ordering has this
property.

Hausdorff classified countable linear orderings according 
to their rank. It follows from results proved in
\cite{Heilbrunner} that the rank of the 
lexicographic ordering of a scattered regular language is always
finite. It is known (cf. \cite{BEscattered}) 
that a CPA $\cA$ accepts a scattered regular language
iff for each nontrivial strongly connected component $C$ 
and $q \in C$, either $q0 \not\in C$ or $q1 \not\in C$.
It would be interesting to know whether there is 
a polynomial time algorithm to decide whether two DFA 
accepting scattered languages accept isomorphic languages.

\paragraph*{Acknowledgement}
The author would like to thank all three referees for suggesting improvements
and Szabolcs Iv\'an for the references on addition chains. 


\section*{Appendix}

Suppose that $\cA = (Q,\zo,\delta,q_1,F)$ with $Q = \{q_1,\ldots,q_n\}$
is a DFA having no nontrivial strongly connected component over 
the binary alphabet $\zo$ such that all final states 
are sinks, i.e., whenever $q$ is a final state, neither $q0$ nor $q1$ 
is defined.

{\bf Algotithm 1} {\em Input}: A word $u$ (of length less than $n$) such that 
neither $q_1u0$ nor $q_1u1$ is defined. 

{\em Output}: The number of different words $v$ accepted by
$\cA$ such that $u \lex v$.

{\em Method}: 
Let $M_0$ and $M_1$ denote the $Q\times Q$ transition
matrices of $\cA$ with respect to the letters $0$ and $1$,
respectively. 
Let $u = u_1\cdots u_k$, where each $u_i$ is either $0$ or $1$.
For each $\ell$ with $1 \leq \ell \leq k$ and $u_\ell = 0$,
consider the sum
$N_\ell = M_{u_1}\cdots M_{u_{\ell -1}}M_1\sum_{j = 1}^{n -\ell}(M_0 + M_1)^j,$
where matrix sum and product are computed in the semiring of natural numbers. 
Since each word of length $n$ or longer induces the 
empty partial function on the set of states, is clear that for each $1 \leq i,j \leq n$, 
$(N_\ell)_{ij}$ is the number of words accepted by $\cA$ with 
initial state $q_i$ and final state $q_j$ of the form 
$u_1\cdots u_{\ell-1}1x$. 

Let $e$ denote the $Q$-dimensional row vector whose first entry is $1$
and whose other entries are $0$, and let $f$ denote the $Q$-dimensional 
$0$-$1$ column vector whose $q_i$th component is $1$ iff $q_i \in F$,
for $1 \leq i \leq n$.  
Then, $\sum_{u_\ell = 0} e N_\ell f$ is the number of all words 
accepted by $\cA$ lexicographically greater than $u$.
By the above consideration, and since each number 
occurring in the computation is at most $2^n$ that 
can be represented by $n+1$ bits, this number can be computed
in polynomial time in the number $n$ of states.

{\bf Algorithm 2}
{\em Input}:  A state $q \in F$.

{\em Output}: The lexicographically greatest word $u$ with $q_1u = q$.

{\em Method}: First, in polynomial time, compute the $Q \times Q$ 
binary reachability matrix $M$ such that $M_{q_i,q_j} = 1$ iff there is a word
$u$ with $q_iu = q_j$. Then form a finite sequence of states 
$s_1,\cdots,s_k$  together with letters $u_1,\cdots,u_{k-1}$
such that $s_1 = q_1$, and if $s_i \neq q$ then 
$s_{i+1} = s_i1$ and $u_i = 1$ if $M_{s_i1,q} = 1$; and $s_{i+1} = s_i0$ 
and $u_i = 0$ otherwise. 
The length $k$ of this sequence is at most $n$ 
and $u_1\cdots u_{k-1}$ is the lexicographically greatest word $u$ 
with $q_1u = q$.

\end{document}